\documentclass[aps,preprint]{revtex4}
\usepackage{epsfig}
\usepackage{amsmath}
\usepackage{epsfig}

\begin{document}
\title
{\bf The paradoxical zero reflection at zero energy}    
\author{Zafar Ahmed$^1$, Vibhu Sharma$^2$, Mayank Sharma$^3$, Ankush Singhal$^4$, Rahul Kaiwart$^5$, Pallavi Priyadarshini$^6$}
\affiliation{$~^1$Nuclear Physics Division, Bhabha Atomic Research Centre, Mumbai 400085, India\\ 
$^{2,3}$Amity Institute of Applied Sciences, Amity University, Noida, UP, 201313, India\\
$^4$Department of Physics, UM-DAE-CBS, Mumbai, 400098, India\\
$^{5,6}$Human Resource Development Division, Bhabha Atomic Research Centre, Mumbai400085, India}
\email{1:zahmed@barc.gov.in, 2: svibhu876@gmail.com, 2: mayank.edu002@gmail.com, 4:ankush.singhal@cbs.ac.in,5:rahul.kaiwart@gmail.com,6:ppriyadarshini@gmail.com}
\date{\today}
\begin{abstract}
\noindent
Usually, the reflection probability $R(E)$ of a particle of  zero  energy incident on a potential which converges to zero asymptotically is found to be 1: $R(0)=1$. But earlier, a paradoxical phenomenon of zero reflection at zero energy ($R(0)=0$) has been revealed as a threshold anomaly. Extending the concept of Half Bound State (HBS) of 3D, here we show that in 1D when a symmetric (asymmetric)  attractive potential well possesses a zero-energy HBS, $R(0)=0$ $(R(0)<<1)$. This can happen only at some critical values $q_c$ of an effective parameter $q$ of the potential well in the limit $E \rightarrow 0^+$. We demonstrate this critical phenomenon in two simple analytically solvable models which are square and exponential wells. However, in numerical calculations  even for these two models $R(0)=0$ is observed only as extrapolation to zero energy from low energies, close to a  precise critical value $q_c$. By numerical investigation of  a variety of potential wells, we conclude that for a given potential well (symmetric or asymmetric), we can adjust the effective parameter $q$ to have a  low reflection at a  low energy. 
\end{abstract}
\maketitle
\section{Introduction}
Usually, the reflection probability $R(E)$ of a particle of  zero (extremely low) energy incident on a one-dimensional potential which converges to zero asymptotically is found to be 1: $R(0)=1$, the single Dirac delta and the square  well potentials are the simplest examples [1-6]. This observation is also intuitive, for a zero-energy particle the tunnel effect is negligible such that the transmission probability is close to zero. Earlier, a paradoxical phenomenon that $R(0)=0$ has been proposed and proved as a threshold anomaly [7] for a potential which is at the threshold of binding a state at $E=0.$ This paradoxical result may be understood in terms of wave packet scattering from an attractive potential. A wave packet with zero average kinetic energy, localized to one side of the potential, will spread in both directions. When the low energy components scatter against the potential, they are transmitted and this would appear simply as wave packet spreading preferentially to the  other direction.
 
Here we show that it is rather a critical effect which occurs when a scattering potential well becomes critical: possesses a Half Bound State (HBS) at $E=0$. We extend the concept of HBS to one dimension. HBS  is  discussed [1-3,8] in low energy scattering from a three dimensional  central potential in terms of scattering length. In two analytically solvable models, we show that both HBS at $E=0$ and $R(0)$ occurs when an effective parameter $q$ of the potential takes a critical discrete value  $q_c$. However in numerical calculation of even these two models, we show that $R(0)=0$ is achieved as an extrapolation from low energies to zero energy that too when $q$ equals $q_c$ very accurately. In this regard, very low reflection at very low energies is no less surprising and we show that it is 
plausible and it could even be more practical than $R(0)=0$.

\begin{figure}[t]
\centering
\includegraphics[width=4 cm,height=5 cm]{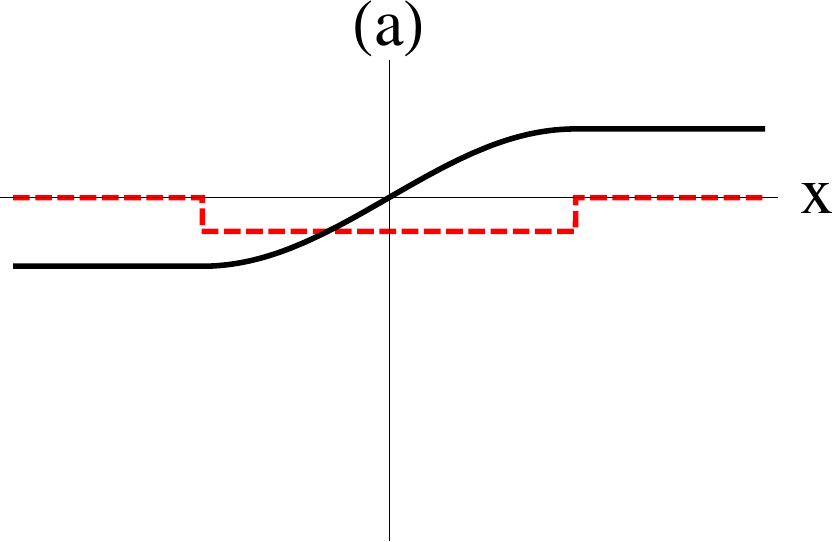}
\hskip .5 cm
\includegraphics[width=4 cm,height=5 cm]{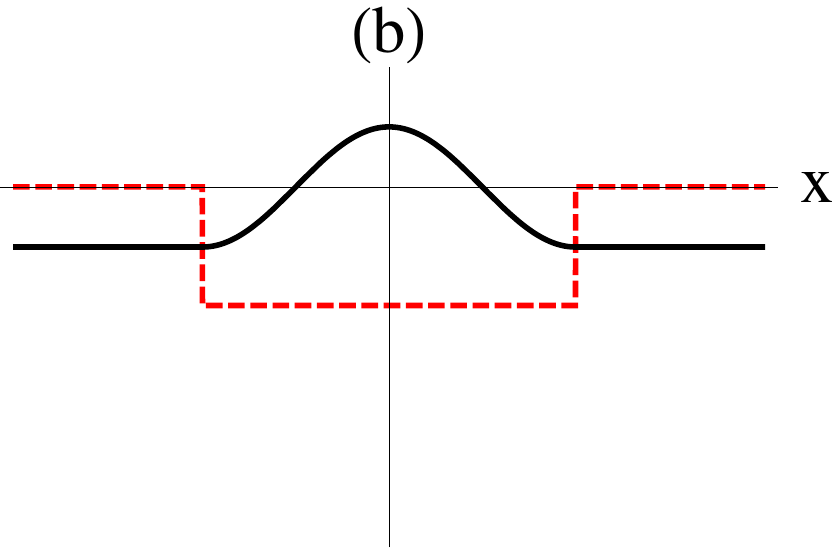}
\hskip .5 cm
\includegraphics[width=4 cm,height=5 cm]{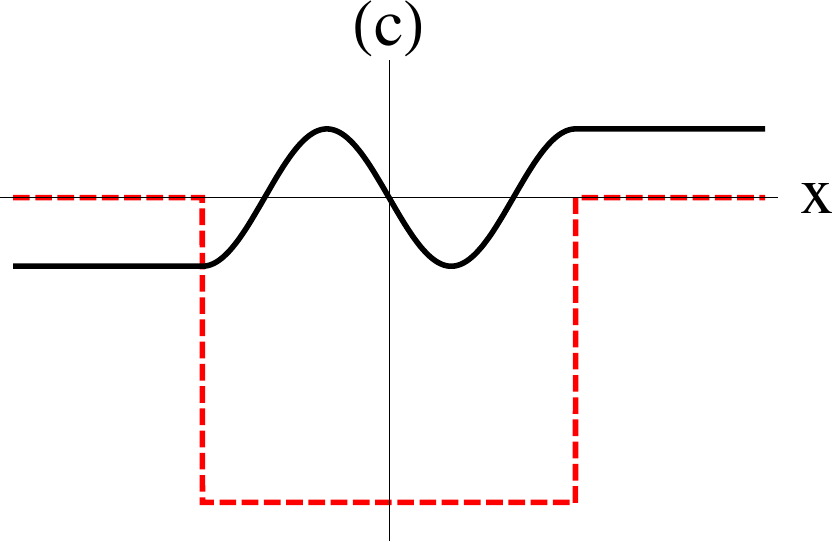}
%\hskip .5 cm
%\includegraphics[width=3 cm,height=5 cm]{fig-pdx-1d.pdf}
\caption{Depiction of half bound state $\psi_*(x)$ (10) (solid lines) at $E=0$ in one dimensional square well potentials (dashed lines)  when their depth is increasing and admitting three critical values in (a,b,c). We have taken $a=1$, $2\mu=1=\hbar^2$, so the depth of the well is $V_0=q_c^2$. Here we consider $q_c=n\pi/2, n=1,2,3$ where in addition to 1 HBS at $E=0$, the wells have 1,2, and 3 bound states in (a,b,c) for $E<0$, respectively. These bound states are not shown here. These critical square well potentials are shown to have $R(0)=0$ in the section II-A.}
\end{figure} 
Three dimensional zero angular momentum (s-wave) Schr{\"o}dinger equation for a central potential $V(r)$ is written as
\begin{equation}
\frac{d^2w(r)}{dr^2}+\frac{2 \mu}{\hbar^2}[E-V(r)]w(r)=0, \quad w(r)=r\psi(r).
\end{equation}
One demands $w(0)=0$ so that the wave function $\psi(r)$ is regular at origin. When $E$ is very small and $V(r)$ vanishes at large distances such that $[E-V(r)] \approx 0$, the solution of (1) for $r\rightarrow \infty$ can be given as $w(r) \sim Ar+B$ and the scattering length [1-3,8] is defined as $a_s=-B/A$. It has been shown [9] that when the depth of a potential is increased, $a_s(V_0)$ varies from positive to negative and {\it vice versa} by becoming discontinuous ($\pm \infty$) at certain discrete values say $V_{00}, V_{01}, V_{02},..$. For  $V_0=V_{0n}$,  $|a_s|$ is very large, then  by increasing the depth $V_0$ slightly the potential can be made to possess a weakly bound state at an energy slightly below $E=0$. So an infinite scattering length is a signature that the potential well possesses a HBS at $E=0$ or it is  at the threshold of possessing one more bound state at $E<0$. Further, $a_s=\pm \infty$ implies $A=0$ and  the wave function becomes constant and parallel to $r$-axis, asymptotically:  $w(r\ge L)=B$. This non-normalizable state is called half bound state [1-3,8] and we  can also  characterize it with the Neumann boundary condition that $w'(L)=0$, where $L$ may be finite for a short-ranged potential  or infinite for a potential that converges to zero asymptotically. As pointed out in [7], Wigner [11] has called such a state as resonance near threshold, Schiff [12] calls it a bound state near continuum which causes resonance in the scattering cross-sections due to a central  potential. Using the attractive exponential potential well: $V(r)=-V_0\exp(-r/a)$, the resonance in scattering cross section has been demonstrated [5] when the strength parameter $\sqrt{8\mu V_0a^2}/\hbar$ coincides with the zeros of the cylindrical Bessel function $J_0(z).$

The one-dimensional time-independent  Schr{\"o}dinger equation is written as
\begin{equation}
\frac{d^2\psi(x)}{dx^2}+\frac{2 \mu}{\hbar^2}[E-V(x)]\psi(x)=0.,
\end{equation}
where $\mu$ is the mass of the particle and $\hbar$ is the Planck constant divided by $2\pi$. Let us define
\begin{equation}
k=\sqrt{\frac{2 \mu E}{\hbar^2}}, E>0, \quad \kappa =\sqrt{\frac{-2 \mu E}{\hbar^2}}, E<0, \quad q=\sqrt{\frac{2 \mu V_0a^2}{\hbar^2}}, V_0>0,
\end{equation}
which are useful in the sequel. Here, $V_0, a$ and $q$ are the depth, the width  and the effective strength parameters of the potential well, respectively. Let $u(x)$ and $v(x)$ be two linearly independent real solutions of (2) such that their wronskian  $W(x)= u(x)v'(x)-u'(x)v(x)$  is constant (position-independent) for all real values of $x$. We may choose $u(0)=1, u'(0)=0; v(0)=0, v'(0)=1$ [3] to start numerical integration of (2) on both sides left and right. Let the scattering attractive potential $V(x)$ be non-symmetric and converging to zero at $x=\pm \infty$. Let $x=-L_2$ and $x=L_1$ be the distances where  $V(x)$ is extremely small. We propose to give the condition for HBS at $E=0$ as
\begin{equation}
\psi^\prime(-L_2)= 0=\psi^\prime(L_1).
\end{equation} 
In contrast to the bound states, HBS do not vanish asymptotically; they instead
saturate to become constant (parallel) there (see Figs., 1,2). A slight increase (decrease) in depth of the well can make this state bound (unbound).
If $V(x)$ is symmetric $(L_1=L_2=L)$, the solutions $u(x)$ and $v(x)$ are of  definite parity (even and odd, respectively). The conditions for HBS are
\begin{equation}
u(0)=1, u'(L)=0 \quad \mbox{or} \quad v(0)=0, v'(L)=0.
\end{equation}
If $V(x)$ is not symmetric, we have HBS at $E=0$ such that
\begin{equation}
u'(-L_2) =0= u'(L_1) \quad {\mbox or} \quad v'(-L_2)= 0 =v'(L_1).
\end{equation}
Imposition of these boundary conditions on the second order differential equation (2) yields the critical values $q_c$ of the effective parameter $q$ (3) of the well for a HBS at $E=0$. A scattering potential well which is such that $\int_{-\infty}^{\infty} V(x) dx < 0 $, has at least one [10] bound state  for howsoever small value of $q$. So a HBS has at least one node. Amusingly the node less  HBS is nothing but constant: $\psi(x)=C$ which exists when the depth of the potential is set equal to zero !  For the symmetric case, $x=0$ is the node and for the non-symmetric case the node could be found at $x=l$, where $-L_2 < l < L_1.$ If at $E=0$ the well has the solitary HBS of ${\cal N}$-nodes, it will have ${\cal N}$ number of bound state for $E<0$.

To illustrate an HBS, one can readily check for $V(x)=-2 ~\mbox{sech}^2 x$ there  is one node less ground state $\psi_0(x)= B ~\mbox{sech} x$ ($\mbox{sech}x=2/[\exp(x)+\exp(-x)])$ at $E=-1$, whereas $\psi_*(x)= A \tanh x$ is a HBS at $E=0$. In one dimension HBS is usually ignored. Henceforth, we propose to denote  HBS  as $\psi_*(x)$ against the notation $\psi_n(x)$ for the bound states. 

\begin{figure}[t]
\centering
\includegraphics[width=4 cm,height=5 cm]{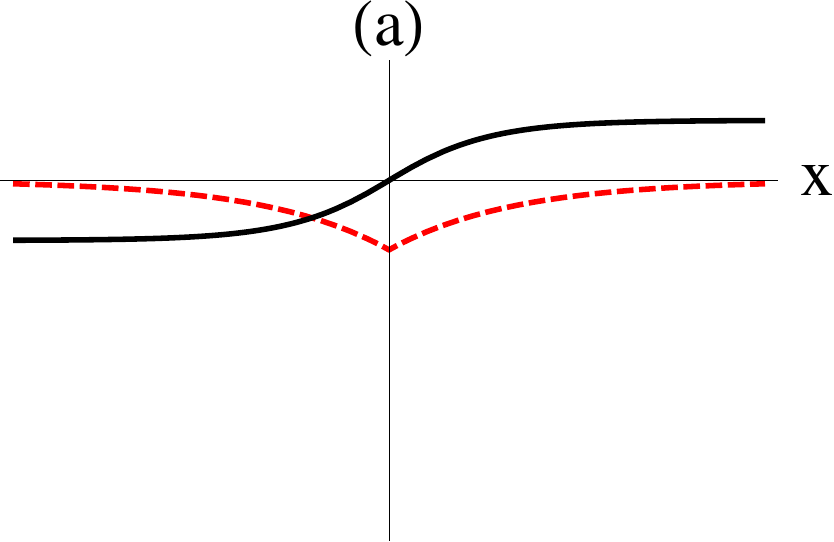}
\hskip .5 cm
\includegraphics[width=4 cm,height=5 cm]{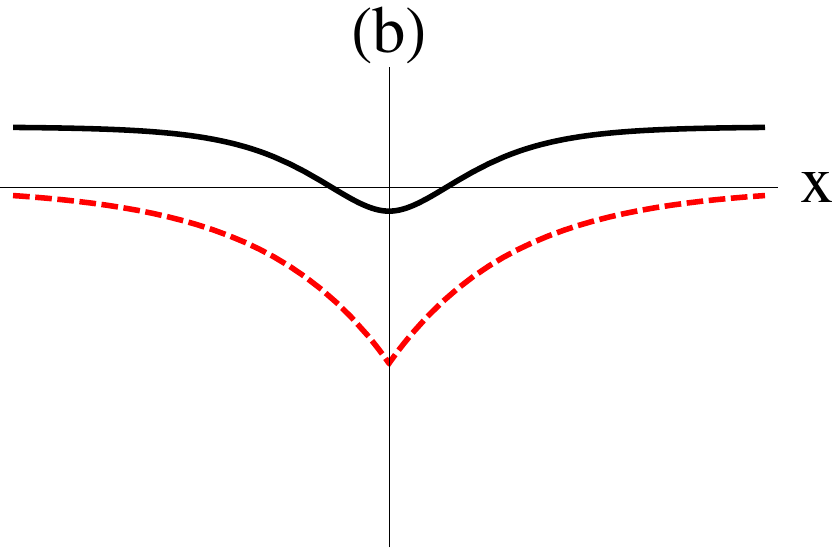}
\hskip .5 cm
\includegraphics[width=4 cm,height=5 cm]{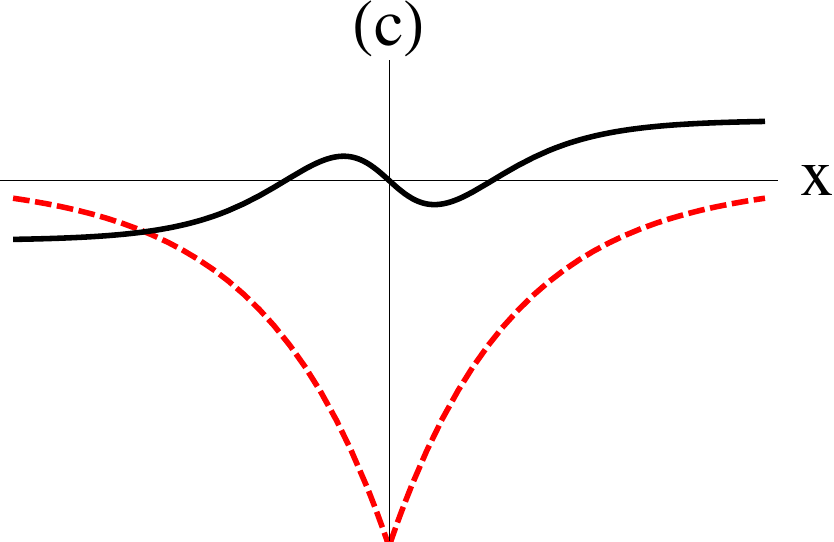}
%\hskip .5 cm
%\includegraphics[width=3 cm,height=5 cm]{fig-pdx-2d.pdf}
\caption{The same as in Fig. 1, $\psi_*(x)$ for the exponential well (11), where $q_c=2.40$ (first zero of $J_0(z)$), 3.83 (First zero of $J_1(z)$) and $5.52$ (second zero of $J_0(z)$).  For these potentials, $R(0)=0$ has been demonstrated in the 
	section II-B}
\end{figure}
Let us denote $u(L_1)=u_1, v(L_1)= v_1, u'(L_1)=u'_1, v'(L_1)= v'_1$
and $u(-L_2)=u_2, v(-L_2)= v_2, u'(-L_2)=u'_2, v'(-L_2)= v'_2.$ Following the Appendix of Senn [7] for reflection amplitude we write
\begin{equation}
r(E)=\frac{B}{A}= -\frac{[u'_2 v'_1 -u'_1 v'_2] +ik [v_2 u'_1+ u_1 v'_2]-ik[u_2 v'_1 +v_1 u'_2]+k^2 [u_1 v_2-u_2 v_1]}{[u'_2 v'_1 -u'_1 v'_2] -ik [v_2 u'_1- u_1 v'_2]+ik[u_2 v'_1-v_1 u'_2]-k^2 [u_1 v_2-u_2 v_1]} e^{-2ika}.
\end{equation}
The reflection probability (factor) is given by $R(E)=|r(E)|^2$.
Ordinarily, when $E=0$, $r(0)= -\frac{u'_2 v'_1 -u'_1 v'_2}{u'_2 v'_1 -u'_1 v'_2}=-1$, provided $v'_1,v'_2 \ne 0$. 
But when at $E=0$ and half the  bound state condition:(5) or(6) is satisfied, $r(0)=0/0$ (indeterminate). In order to find limit of $r(E)$ as $E \rightarrow 0^+$, in Eq. (7),  we can first set  $[u'_2 v'_1 -u'_1 v'_2] =0$ due to the HBS connection (5,6), cancel $k$, then using $u'_1=0=u'_2$ and $v'_1=0=v'_2$ (5,6) again, one finds
$\lim_{E\rightarrow 0} r(0)= \frac{u_1 v'_2-u_2 v'_1}{u_1 v'_2 + u_2 v'_1}$, \quad
$\lim_{E\rightarrow 0} r(0)= \frac{v_2 u'_1-v_1 u'_2}{v_2 u'_1+v_1 u'_2}$,
respectively. Eventually, when $V(-x)=V(x)$, $u(x)$
and $v(x)$ acquire definite parity (even and odd, respectively) and we have $u_1=u_2, v_1=-v_2; u'_1=-u'_2, v'_1=v'_2$ yielding $R(0)=0$ [7]. This completes our rephrasing of zero reflection at zero energy when an attractive well possesses a half bound state at zero energy.

We find that the single Dirac delta well potential [3] in any case yields $R(0)=1$ and becomes a trivial exception to the zero reflection at zero energy. In section II, we present two illustrations of attractive potentials
possessing zero energy bound state an $R(0)=0$. In section III, we explore
low reflection at a low energy in various attractive potential wells

\section{Illustrations: $R(0)=0$ and HBS at zero energy}
\subsection{Square well potential:}
The most common square well potential is given as  $V(-a<x<a)=-V_0, V(x)= 0$ (otherwise) its reflection factor is written  as [1-6]
\begin{equation}
R(E)=\frac{\sin^2 2q \sqrt{1+\epsilon}}{4\epsilon(\epsilon+1)+\sin^2 2q\sqrt{1+\epsilon}},\quad \epsilon=E/V_0.
\end{equation}
Ordinarily, $R(0)=\frac{\sin^2 2q}{\sin^2 2q}=1$, unless and until $q=n\pi/2=q_c, n= 1,2,3...$
It is in these special cases that $R(0)$ becomes indeterminate $(0/0)$ and then one has to take  $\lim_{E \rightarrow 0^+} R(E)$ properly by L'Hospital rule (see [21]): where differentiation of the numerator and the denominator  with respect to $E$ (separately) yields
\begin{equation}
\lim_{E \rightarrow 0^+} R(E)= \lim_{E \rightarrow 0^+}\frac{n\pi \sin ( 2 n \pi \sqrt{1+\epsilon})}{2\sqrt{1+\epsilon} (8 \epsilon +4)+ n \pi \sin (2 n \pi \sqrt{1+\epsilon})}=0, \quad n = 0,1,2,3,...
\end{equation}
For $E=0$, the solution of Schr{\"o}dinger equation can be given as
\begin{eqnarray}
\psi_*(x)&=& \left\{\begin{array}{lcr}
A~\sin \frac{n\pi x}{2a}, \quad |x|<a \\ 
A~\mbox{sgn}(x)~\sin (n \pi/2),  \quad |x| \ge a, n\mbox{(odd)}
\end{array}
\right. \nonumber \\ 
\psi_*(x)&=& \left\{\begin{array}{lcr}
A~\cos \frac{n\pi x}{2a}, \quad |x|<a \\ 
A~\cos (n \pi/2),  \quad |x| \ge a, n\mbox{(even)}
\end{array}
\right. 
\end{eqnarray}
where $\mbox{sgn}(x)= -1, x<0, \mbox{sgn}(x)= +1, x>0$. So $\psi_*(x)$ is a HBS satisfying the conditions (5), here $L=a$. In Fig.1 we plot first three $(E=0)$ HBS for $q_c=n\pi/2, n=1,2,3$. This potential  may be dismissed  to be a very special one, for instance it has energy oscillations in $R(E)$. So below, we present the exponential potential as a nontrivial example.
\subsection{The exponential potential well}
This symmetric attractive potential which vanishes asymptotically is expressed as
\begin{equation}
V(x)=-V_0 \exp({-2|x|/a)}, \quad a, V_0>0.
\end{equation}
The exponential potential is also a commonly discussed central potential for both bound and scattering states [3,5]. Its reflection amplitude is given in terms of the cylindrical Bessel functions $J_\nu(z)$ [13] as [6]
\begin{equation}
r(E)=-\frac{1}{2} \left(\frac{q}{2}\right)^{-2ika} \frac{\Gamma(1+ika)}{\Gamma(1-ika)} \left ( \frac{J_{ika}(q)}{J_{-ika}(q)} + \frac{J'_{ika}(q)}{J'_{-ika}(q)}, \right).
\end{equation}
Here $\Gamma(z)=\int_{0}^{\infty} x^{z-1} \exp(-zx) dx, {\cal R}(z)>0$ [13].
It may be readily checked that the limit of $r(E)$ as $E \rightarrow 0^+$ is -1, until $q$ is a zero of the function $J_0(z)$. In this case $r(0)=0/0$ is indeterminate. In order to get to the correct limit one can Maclaurian expand $J_{\nu}(z)$ about $\nu=0$ so for very small values of  $\nu$, one can write $J_{\nu}(z) \approx J_{0}(z)+ \frac{\nu \pi}{2} Y_{0}(z)$ [13], where $Y_0(z)$ is zeroth order Neumann function. We also use a result that $J'_0(z)=-J_1(z), Y'_0(z)=-Y_1(z)$ [13]. So for  values of $E \rightarrow 0^+$ we can write                                                                                                                                                                                                                                                                                                                                                                                                                                                                                                                                                                                                                                                                                                                                                                                                                                                                                                                                                                                                                                                                                                                                                                                                                                                                                                                                                                                                                                                                                                                                                                                                                                                                                                                                                                                                                                                                                                                                                                                                                                                                                                                                                                                                                                                                                                                                                                                                                                                                                                                                                                                                                                                                                                                                                                                                                                                                                                                                                                                                                                                                                                                                                                                                                                                                                                                                                                                                                                                                                                                                                                                                                                                                                                           
\begin{equation}
r(E)=-\frac{1}{2} \left (\frac{J_0(q)+ika Y_0(q)}{J_0(q)-ika Y_0(q)} + \frac{J_1(q)+ikaY_1(q)}{J_1(q)-ikaY_1(q)} \right), \quad  E \sim 0.
\end{equation}
Clearly  if $J_0(q)=0$ or $J_1(q)=0$ i.e. $q$ coincides with the well known [13] zeros of $J_0(z)$, and $J_1(q)$; $r(0)=0$. Therefore, the critical values $q=q_c$ are the zeros of the cylindrical Bessel functions $J_0$ and $J_1$.

For bound states, let us insert (11) in (2), the two linearly independent solutions are well known [3,5,6] as $\psi(x)= J_{\pm \kappa a}(q \exp(-|x|/a))$.
For very small values of $z$, $J_{\nu}(z) \approx \frac{(z/2)^{\nu}}{\Gamma(1+\nu)}.$ So we note that choosing $J_{\kappa a}(z)$, we get $\psi(x\sim \infty) \sim \exp(-\kappa x )$  and $\psi(x \sim -\infty) \sim \exp(\kappa x)$ the correct asymptotic behaviour of bound states. Since the potential is symmetric, we can choose two linearly independent solutions $u(x)$ and $v(x)$ which are of even and odd parity respectively such that $u(0)=C_1,u'(0)=0; v(0)=0, v'(0)=C_2$. For even parity states we write
\begin{equation}
u(x)=A~ J_{\kappa a}(q\exp({-|x|/a)}), \quad J'_{\kappa a}(q)=0 \quad (J_{\kappa a}(q) \ne 0).
\end{equation}
For odd parity states we write
\begin{equation}
v(x)=\mbox{sgn}(x)~B~ J_{\kappa a}(q\exp({-|x|/a)}),\quad J_{\kappa a}(q)=0 \quad (J'_{\kappa a}(q) \ne 0),
\end{equation}
where $\mbox{sgn}(x)=-1, x<0 ; ~ \mbox{sgn}(x)= 1, x>0.$
Notice that in both the cases conditions of continuity and differentiability of the eigenstates are satisfied under the given eigenvalue conditions. For fixed value of $q$ the equations
\begin{equation}
J'_{\kappa _na}(q)=0, \quad J_{\kappa _na}(q)=0, \quad \kappa _n=\sqrt{\frac{-2\mu E_n}{\hbar^2}}
\end{equation}
yield the eigenvalues $E_n$ of even and odd parity eigenstates, respectively. In the reflection amplitude (13), if we replace $k$ by $i\kappa _n$, it is instructive to check that the Eqs. (15,16) represent the  negative energy physical poles  of $r(E)$. The bound state eigenvalues are the common poles of the reflection and transmission amplitudes [15] of a one-dimensional potential well. 

From the solutions (15,16) we can identify the zero-energy HBS as 
of odd and even parity
\begin{eqnarray}
\psi_*(x)&=&A~ \mbox{sgn}(x) J_0(q \exp({-|x|/a)}), \quad \mbox{when} \quad \psi_*(0)=J_0(q)=0,\quad \mbox{and} \nonumber \\
\psi_*(x)&=&B J_0(q \exp({-|x|/a)}),\quad  \mbox{when} \quad \psi_*(0) \ne 0, \quad J'_0(q)=0,
\end{eqnarray}
respectively.

Further, we suggest that one can {\it now} study at least two more examples: (i) Soliton  potential $V_S(x)=-\nu(\nu -1) \mbox{sech}^2 x$ [3,4,6]  which is  known to be reflectionless for all positive energies: $R(E)=\sin^2 \nu \pi/(\sin^2 \nu \pi +\sinh^2\pi k)$ whenever $\nu=2,3,4..$, we would like to point out that at these values of $\nu$ these potentials have a half bound-state at $E=0$ (with number of nodes 1,2,3,..., respectively)similar to the ones plotted in Figs. 1, 2 and consequently $\lim_{E \rightarrow 0^+} R(E)=0$ (see [21] again) can be found to exist there.  We would like to remark the  HBS usually goes unmentioned in the literature even for a solvable  potential [3,4,6]. (ii) Ginocchio's [14] potential is an advanced level versatile two parameter $(\nu, \lambda)$ extension of $V_S(x)$ which may now be checked to have $E=0$ as a HBS and $R(0)=0$, whenever $\nu= 2,3,4,...$. Here too the  HBS have number of nodes as 1, 2, 3,..., respectively. 
\section{Low reflection at low energies}

\begin{figure}[t]
\centering
\includegraphics[width=7 cm,height=4 cm]{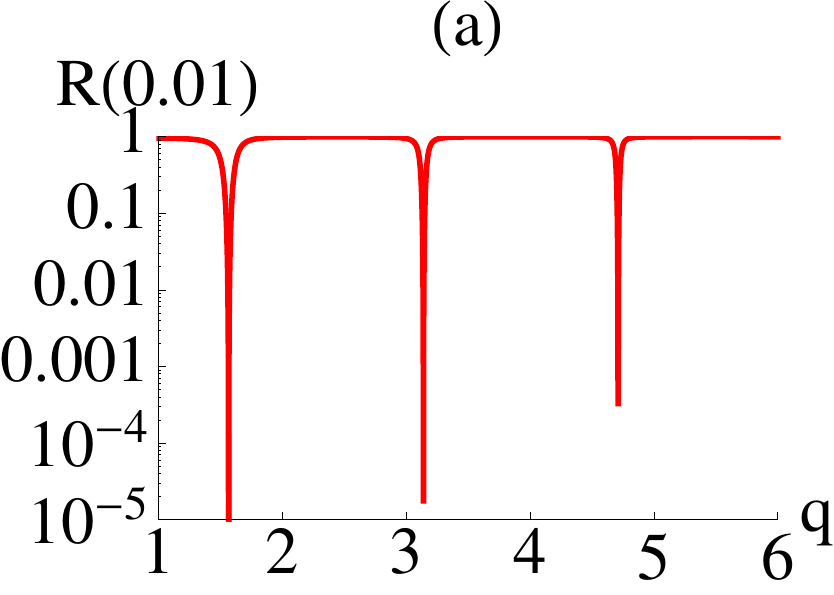}
\hskip .5 cm
\includegraphics[width=7 cm,height=4 cm]{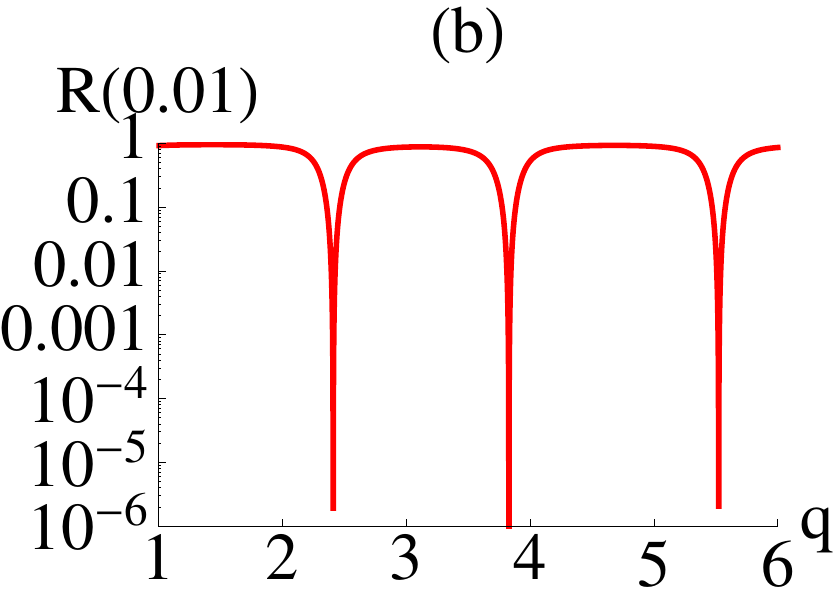}
\caption{Taking $2\mu/\hbar^2=1 (eV {A^0}^{2})^{-1}$, $q=\sqrt{V_0}$,
we plot reflectivity at $E=0.01 eV$ namely $R(0.01)$ as a function of $q$ to show very low or zero-reflection at a
very low energy. (a): for square well when $q$ is in the vicinity of $\pi/2, \pi, 3 \pi/2$, (b): for the exponential well when  $q$ is slightly around $ 2.40, 3.83, 5.52$ (first zero of  $J_0(z)$, first zero of $J_1(z)$, second zero of $J_0(z)$).}
\end{figure}
In the scattering from  two-piece semi-infinite step-barrier potentials which are such that $[V(-\infty)=-V_0, V(\infty)=0, V_0>0$], an interesting existence of a parameter dependent single deep minimum  in reflectivity $R(E)$ at a very low energy  has been revealed [16,17]. However, it seems much earlier [18,19], very low reflection of electrons of very low energies has been measured when electrons cross  a semi-infinite surface (step) barrier. So we understand that the result $R(0)=0$ for the attractive wells could be observed similarly.

The models discussed above in the section II(A,B) are analytically tractable so finding the limit of $R(E)$ as $E \rightarrow 0^+$ is plausible. For practical investigation one would like to actually know the possibility of  low reflection at a low energy around the critical $q$ values of the models of square and exponential wells discussed above. In all the calculations, we shall be using $2\mu =1= \hbar^2$, where $E$ and $a$ in arbitrary units. This choice also means that the mass of the particle  is roughly 4 times of the mass of electron $(\mu=4 m_e)$, wherein mass and energies are measured in electron volt $(eV)$ and lengths  in  Angstrom $(A^0)$ so we have $2\mu/\hbar^2 = 1 (eV {A^0}^2)^{-1}$. In Fig. 3, we plot $R(E=0.01)$ notice extremely low reflectivity around the critical values $q=q_c$ (obtained analytically for $R(0)$ in II(A,B)).

Next important point is to know the behaviour of $R(E)$ when we approach a critical value of the effective parameter $q$ for instance the first zero of $J_0(q)=0$ which is 2.4048255... In Table I, we present this scenario and find that when we are approaching so accurate a value of $q=2.4048255$, we get $R(10^{-5})= 0.1920 \times 10^{-7}$. The Table I, displays a  very slow convergence (numerically) to the result $R(0)=0$, though, this limit has been shown analytically in Eq.(17). For several attractive potential wells, we have used Eq. (7) and an  interesting Matlab recipe [20] for quantum propagation  in 1D systems  based on the method of transfer matrices. We to conclude that
a numerical method of obtaining $R(E)$, will attain $R(0)=0$ as an extrapolation  from low energies to $E=0$. Further, Fig. 3 for the square and the exponential wells  indicates that slightly around the critical value of $q=q_c$ (where $R(0)=0$) one can find  low reflectivity at a low energy $(E=0.01 eV)$. 

In Fig. 4, we present a numerically solved model of low reflection at a low energy. We use the recipe [19] for the numerical computation of $R(E)$ at a low energy $(E=0.1 eV)$  for the case of the parabolic well : $V(x<-a)=0,V(-a<x<0)=V_0(1-x^2/a^2), V(0<x<b)=V_0(1-x^2/b^2), V(x>b)=0.$ See in Fig. 4, (a) for the symmetric case we find $R(0.1)=10^{-6}$ when $q=2.24$, (b) for the asymmetric case $R(0.1)$ is less than $10^{-3}$ when $q=2.13$,  notice that in symmetric case the reflection is much less than that of asymmetric case for the fixed low energy $E=0.1 eV$.

We consider a family of potential wells: $V_{\alpha}(|x|\ge a)=0, V_{\alpha}(|x|\le a)=-V_0[1+\alpha (x-a)/(2a)]$, which change from symmetric square well to an asymmetric triangular well as $\alpha$ varies from 0 to 1. For the case $\alpha=0$, we have the square well which in Fig. 3(a) ($a=1 A^0$) already shows critical values of $q_c$ at or around which $R(0.01)$ is very low. The case of thick but asymmetric well (Fig. 5(a)) sustains the similar characteristics but with higher minima. The more asymmetric case of triangular well ($\alpha= 1$) in Fig. 5(b) does display low reflection around the critical values $q_c$  but these minima in $R(0.01)$ become larger than those in the cases of $\alpha=0, 0.5$ suggesting again that symmetry of a well favours the phenomenon of low reflection at a low energy more. 

Originally, $R(0)=0$ was demonstrated using attractive double Dirac delta well [7] which was a double well potential with extremely thin wells. It is therefore interesting to check whether attractive double wells and multiple wells would preserve the low reflection at low energy. In this regard, we investigate two potentials of finite support such that $V(|x|\ge a)=0$ commonly and  $V_1(|x|<a)=-V_0 \sin^2(\pi x/a)$, $V_2(|x|<a)=-V_0 \sin^2(2\pi x/a)$  (see the dashed lines in the  inset of Fig. 6). These being symmetric wells, in Fig. 6, we confirm very low reflection for  $E=0.01 eV$ at the critical values  $q=q_c$.

\begin{figure}[t]
\centering
\includegraphics[width=7 cm,height=4 cm]{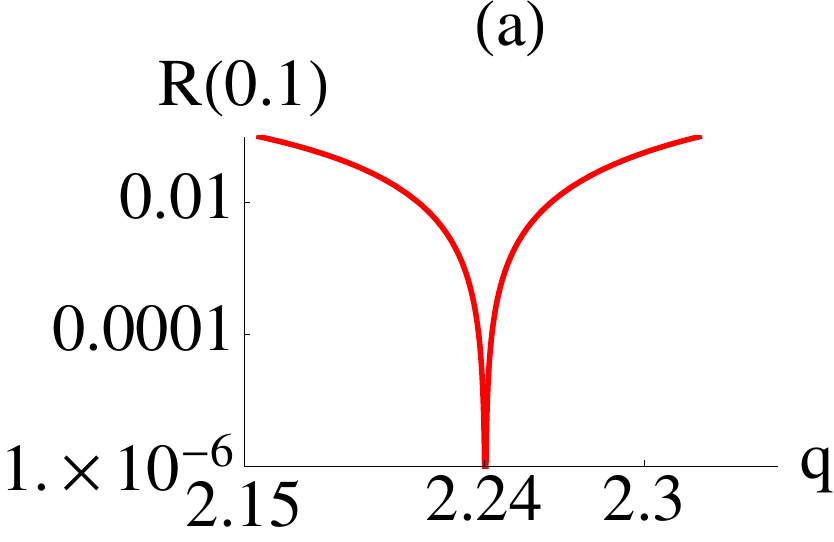}
\hskip .5 cm
\includegraphics[width=7 cm,height=4 cm]{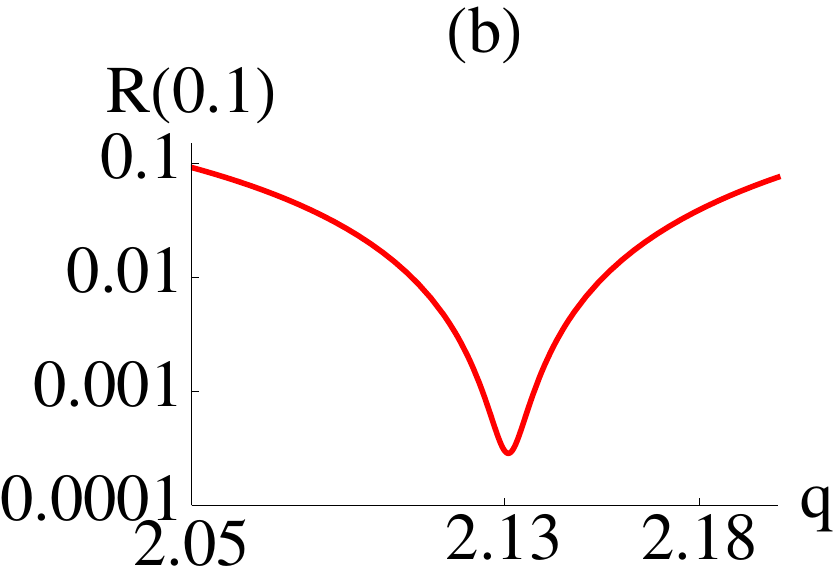}
\caption{
We plot reflectivity at $E=0.1 eV$ namely $R(0.1)$ as a function of $q$ to of the parabolic well show very small reflection for the (a): symmetric case  $(a=1 A^0=b)$ when $q=2.24$ and (b) asymmetric case ($a=1 A^0, b=1.1 A^0$)  when $q=2.13$}
\end{figure}
\begin{figure}[ht]
\centering
%\includegraphics[width=7.5 cm,height=4 cm]{fig-pdx-5a.pdf}
%\hskip .5 cm
\includegraphics[width=7. cm,height=4 cm]{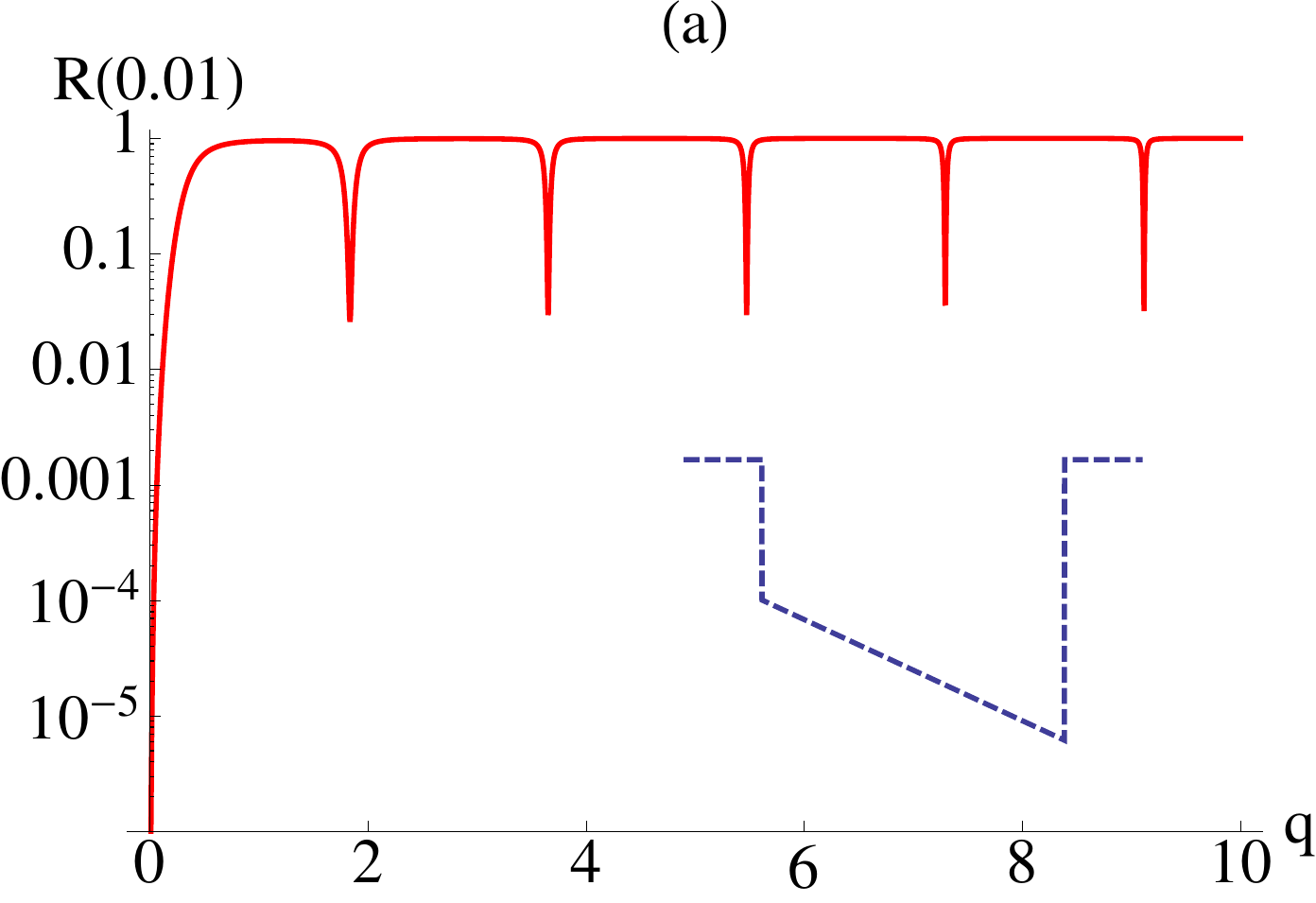}
\hskip .5 cm
\includegraphics[width=7. cm,height=4 cm]{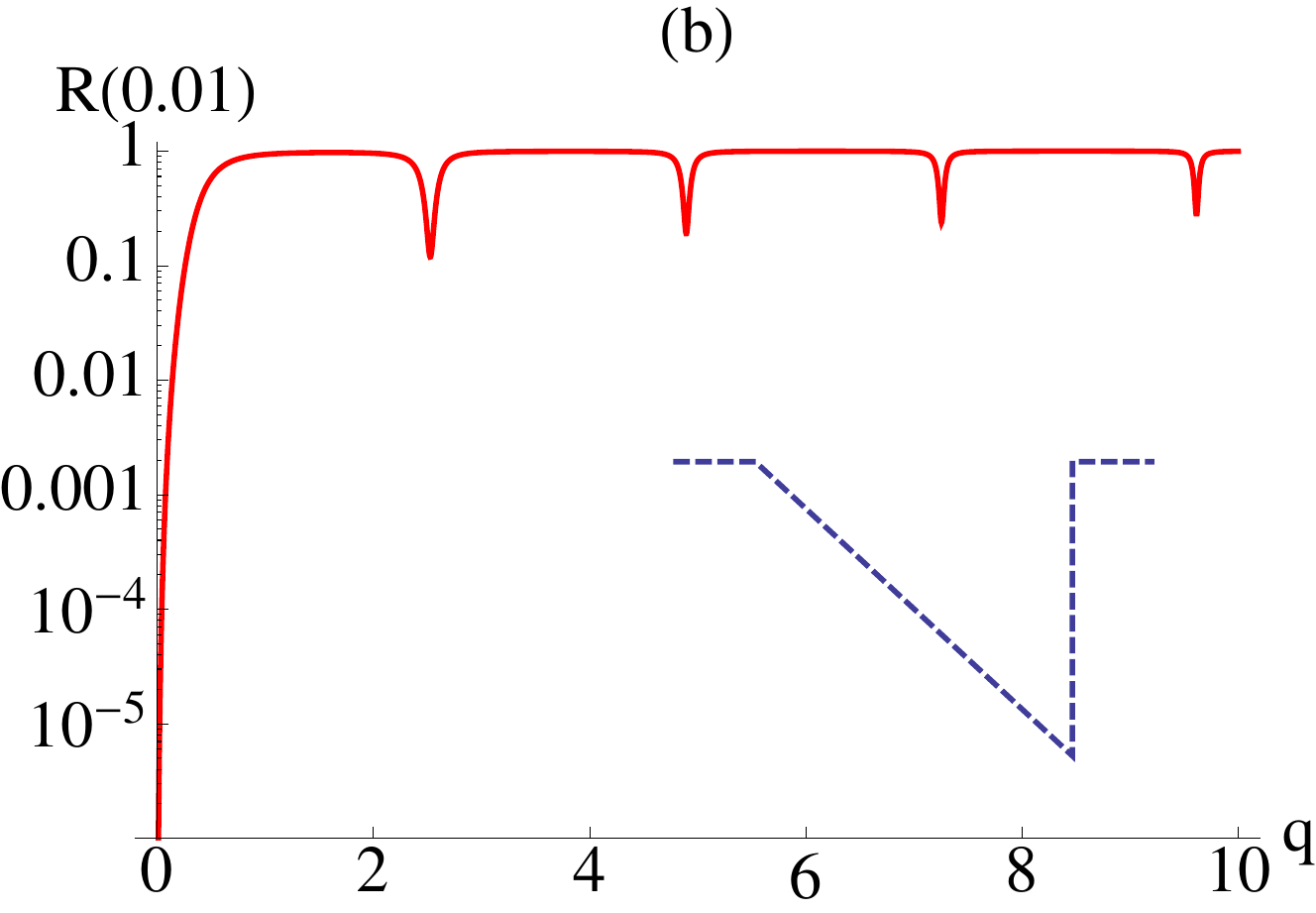}
\caption{$R(E=0.01)$ as a function of $q$ for the square-triangular potential well $V_\alpha(x)$  of depth $V_0=q^2$ (a):  $\alpha=0.50$, (b): $\alpha=1$. For $\alpha=0$, see Fig. 3(a).
}
\end{figure}

\begin{figure}[ht]
\centering
\includegraphics[width=7.5 cm,height=4 cm]{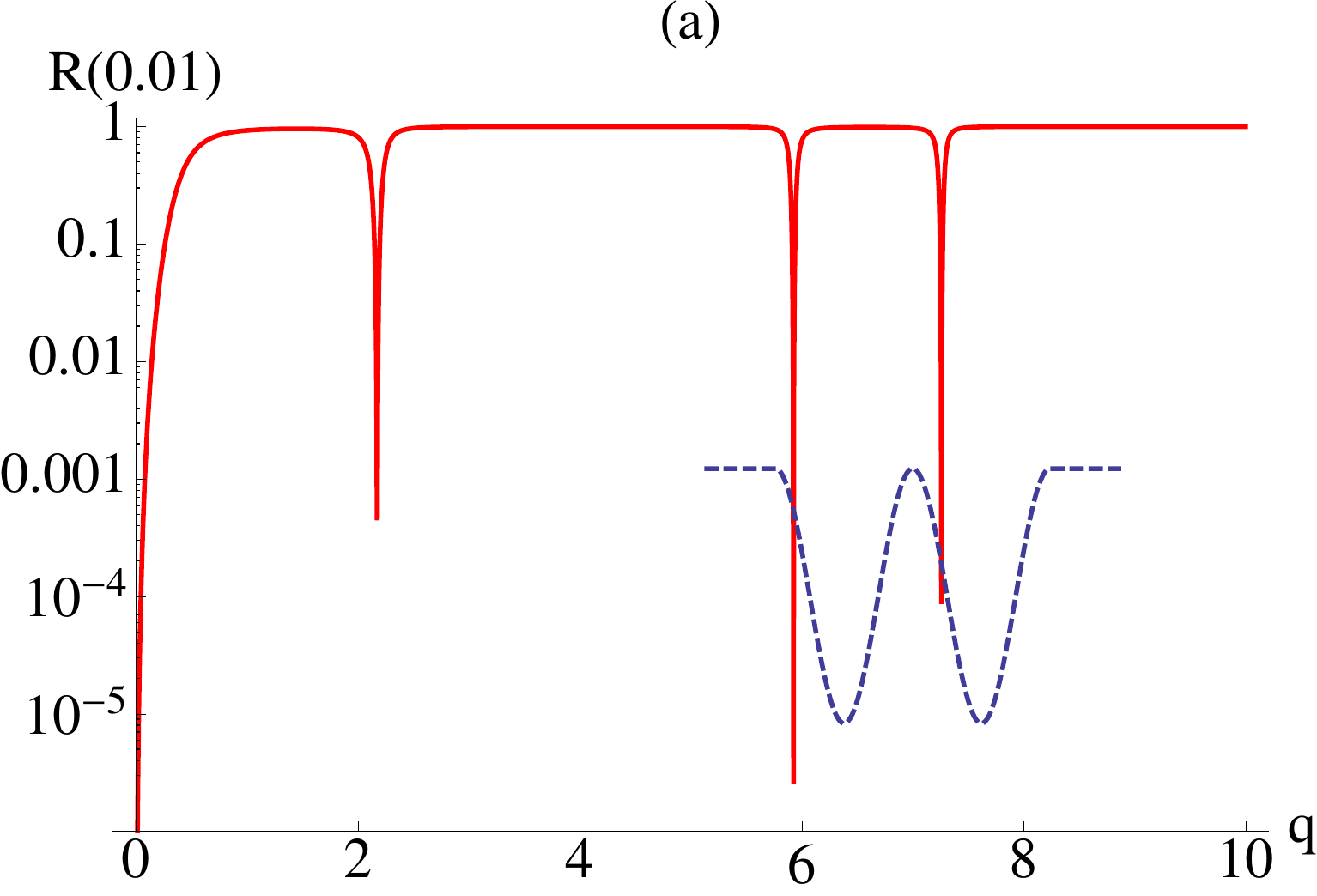}
\hskip .5 cm
\includegraphics[width=7.5 cm,height=4 cm]{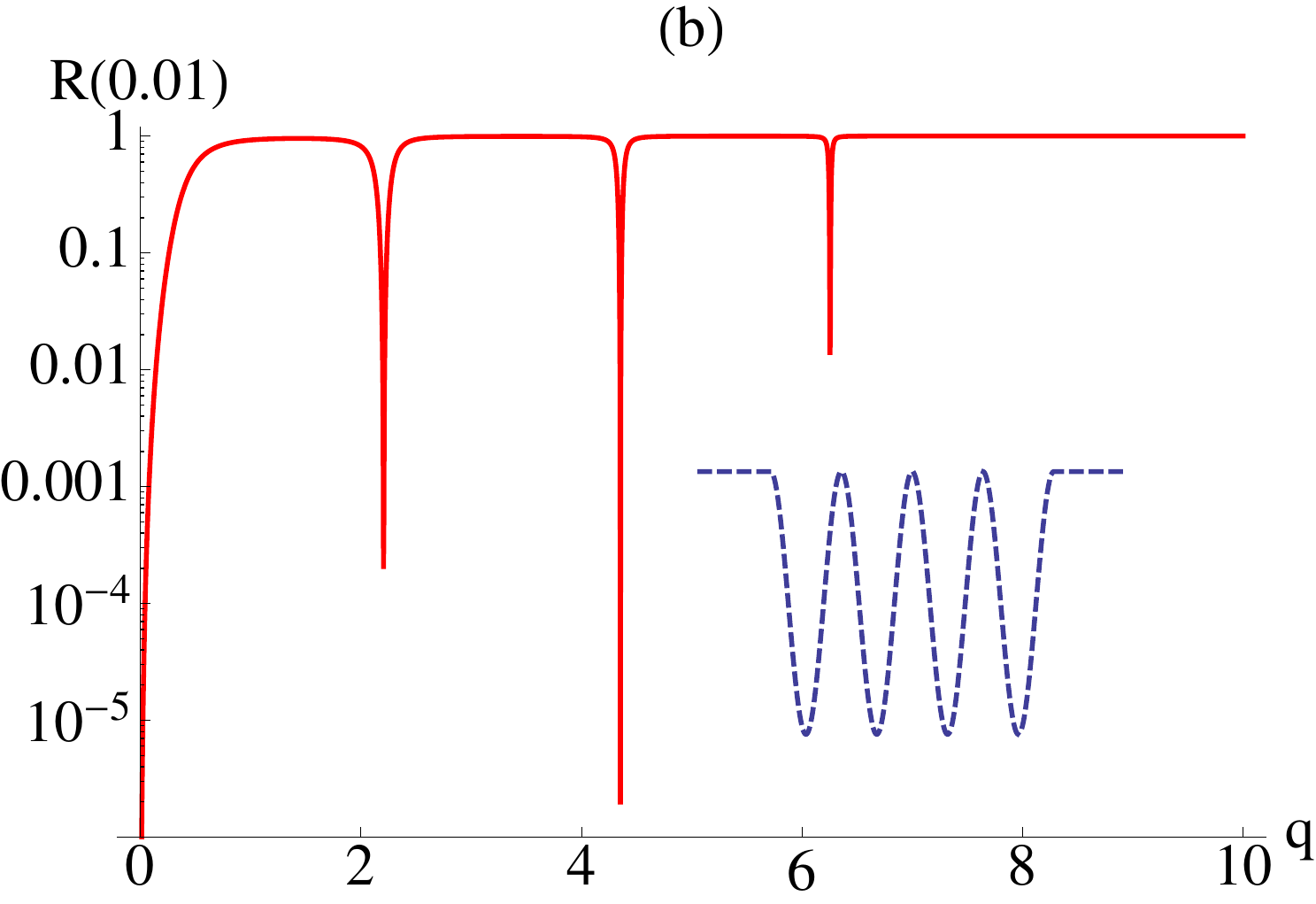}
%\hskip .5 cm
%\includegraphics[width=7 cm,height=4 cm]{fig-pdx-6c.pdf}
\caption{$R(E=0.01 eV)$ as a function of $q$ for the multiple well potentials  $V(|x|>a)=0$: $V_1(|x|<a)=-V_0 \sin^2(\pi x/a)$ and $V_2(x)=-V_0\sin^2(2 \pi x/a)$   in (a) and (b) respectively. The depth parameter $V_0=q^2.$}
\end{figure}

\section{Conclusion}
By extending the concept of zero energy half bound state (HBS) from 3D to 1D,
we have re-phrased the phenomena of $R(0)=0$ ($R(0)<<1$) for the symmetric 
(non-symmetric) attractive potential wells. We hope that this will be found both interesting and instructive. We  denote HBS as $\psi_*(x)$ in distinction to the bound states $\psi_n(x)$. In a scattering potential well (s.t $V(\pm \infty)=0$), the solitary HBS is characterized by Neumann boundary condition that $\psi_*'(\pm L)=0$ (L may be finite or infinite, depending upon whether the well is short ranged or converging to zero asymptotically).  A well having a HBS of ${\cal N}$-nodes  at $E=0$ means that it has (${\cal N}$) number of bound states below $E=0$. A HBS which occurs only at certain critical values $q_c$ of  strength parameter $q$ (3) of the well with one or more number of nodes, is often ignored. We have shown that for a symmetric scattering potential well, zero reflection at zero energy occurs critically at $q=q_c$ and as a limit of $R(E)$ as $E \rightarrow 0^+$ (see Table I). For $R(0)=0$ and its connection with HBS, we have presented two analytic  illustrations of square and exponential wells and suggested two more. We have noted that the single Dirac delta potential which is devoid of HBS and has only one bound state is a trivial exception to this paradoxical phenomenon. However, we believe that it is the low reflection at a low energy which is practically more desirable. In this regard, by investigating several profiles of scattering potential wells we find that for a fixed small energy ($\epsilon$), there exist critical values $q_c$ at or around which the  reflection $R(\epsilon)$ is very small. So we can adjust the strength parameter of a well to get a low reflection at a low energy. The low reflection at a fixed low energy could be much less in case of symmetric wells than  in  asymmetric ones (see Figs. 4,5).

\begin{table}[t]
\caption {The scenario of very low reflectivity at low energies when we approach the critical value of the effective parameter $q=2.4048255...$ obtained analytically (16) for the exponential well (11).}
\begin{tabular}{|c|c|c|c|c|c|}
\hline
$q$ & $R(10^{-1})$ & $R(10^{-2})$ & $R(10^{-3})$ & $R(10^{-4})$  & $R(10^{-5})$ \\
\hline
2.40 & $.1695\times 10^{-1}$    &  $.5423 \times 10^{-2}$  & $.1251 \times 10^{-1}$ & $.9150 \times 10^{-1}$ & $.4956 \times 10^{0}$\\
2.404 & $.1517 \times 10^{-1}$ & $.2320 \times 10^{-2}$ & $.9370 \times 10^{-3}$ & $.3335 \times 10^{-2}$ & $.2828 \times 10^{-1}$\\ 
2.4048 & $.1482 \times 10^{-1}$ & $.1854 \times 10^{-2}$ & $.2029 \times 10^{-3}$  & $.3601 \times 10^{-4}$ &  $.4372 \times 10^{-4}$\\
2.40482 & $.1481 \times 10^{-1}$ &  $.1843 \times 10^{-2}$ & $.1914 \times 10^{-3}$ &
$.2213 \times 10^{-4}$ & $.6316 \times 10^{-7}$\\
2.404825 & $.1481 \times 10^{-1}$ & $.1840 \times 10^{-2}$ &  $.1886 \times 10^{-3}$ & $.1919 \times 10^{-4}$ & $ .2215 \times 10^{-7}$\\
2.4048255 & $.1481 \times 10^{-1}$ & $.1840 \times 10^{-2} $ & $.1883 \times 10^{-3}$ &$ .1890 \times 10^{-4}$ & $.1920 \times 10^{-7}$\\
\hline
\end{tabular}
\end{table}
\section{Acknowledgment} ZA would like  to thank physics trainee-officers of 60$^{th}$ batch of HRDD of BARC for their interest in the course of Quantum Mechanics.


\begin{thebibliography}{37}
\bibitem{21} Let $f(x,y)=y/(y+x)$, then  if $y \ne 0$, $f(0,y)=1$ and also the limit of $f(x,y)$ as $x \rightarrow 0$ is 1. But if $y=0$, $f(0,0)=0/0$ (undefined or indeterminate), then $\lim_{x\rightarrow 0^+} f(x,0)= \lim_{h\rightarrow 0} f(h,0)=0/1=0$
by differentiating $w.r.t.~ x$ or $h$ separately in numerator and denominator as per L'Hospital rule. The same is true of $g(x,y)=y/(y+x^2)$, $g(0,0)=0/0$ is not defined but by L'Hospital rule $\lim_{x \rightarrow 0} g(x,0)= \lim_{h \rightarrow 0} g(h,0) = 0/(2h)=0$ because $h$ is arbitrarily small but not zero or else one can again apply L'Hospital rule.
\end{thebibliography}
\end{document}